\documentclass[11pt,preprint]{aastex}

\slugcomment{Accepted by the Astrophysical Journal, 24 Sept. 2004}

\shortauthors{Boboltz \& Wittkowski}
\shorttitle{Joint VLBA/VLTI Observations of S~Ori}

\begin{document}

\title{Joint VLBA/VLTI Observations of the Mira Variable S~Orionis}
\author{David A. Boboltz}
\affil{U.S. Naval Observatory, \\ 
3450 Massachusetts Ave., NW, Washington, DC 20392-5420, USA \\
dboboltz@usno.navy.mil}
\and
\author{Markus Wittkowski}
\affil{European Southern Observatory, \\ 
Karl Schwarzschild-Str. 2, D-85748 Garching bei M\"unchen, Germany \\
mwittkow@eso.org}

\begin{abstract}
We present the first coordinated VLBA/VLTI measurements of the stellar 
diameter and circumstellar atmosphere of a Mira variable star.
Observations of the $v=1, J=1-0$ (43.1 GHz) and $v=2, J=1-0$ (42.8 GHz) 
SiO maser emission toward the Mira variable S Ori were conducted 
using the VLBA.  Coordinated near-infrared $K$-band measurements of the 
stellar diameter were performed using VLTI/VINCI closely spaced in time
to the VLBA observations.  Analysis of the SiO maser data recorded at 
a visual variability phase 0.73 show the average distance of the masers
from the center of the distribution to be 9.4~mas for the $v=1$ 
masers and 8.8~mas for the $v=2$ masers. The velocity structure of the 
SiO masers appears to be random with no significant indication of 
global expansion/infall or rotation.
The determined near-infrared, $K$-band, uniform disk (UD) diameters 
decreased from $\sim$\,10.5\,mas at phase 0.80 to $\sim$10.2\,mas at 
phase 0.95.  For the epoch of our VLBA measurements,
an extrapolated UD diameter of $\Theta_\mathrm{UD}^K=10.8 \pm 0.3$\,mas
was obtained, corresponding to a linear radius 
of $R_\mathrm{UD}^K = 2.3 \pm 0.5$~AU or
$R_\mathrm{UD}^K =490 \pm 115~R_\odot$.
Our coordinated VLBA/VLTI measurements show that the masers lie 
relatively close to the stellar photosphere at a distance of $\sim$\,2
photospheric radii, consistent with model estimates.  This result is virtually free of 
the usual uncertainty inherent in comparing observations of variable stars 
widely separated in time and stellar phase.  
\end{abstract}


\section{INTRODUCTION}

The evolution of cool luminous stars, including Mira variables, 
is accompanied by significant mass loss to the circumstellar 
envelope (CSE) with mass-loss rates of up 
to $10^{-4}$\,M$_\odot$/year \citep[e.g.][]{JK:90}.
There are currently a number of tools that can be used to study the 
stellar surface and the CSE at various wavelengths.
Optical/near-infrared long-baseline interferometry has provided 
information regarding the stellar diameter, effective temperature, 
center-to-limb intensity variation, and the dependence of these parameters 
on wavelength and variability phase for a number of Mira variables 
\citep[e.g.,][]{HST:95,BDBL:96,PFR:99,YOUNGETAL:00,TCA:02,TCB:02,WED:04}.
The structure and dynamics of the CSE of Mira variables and supergiants 
have been investigated by mapping the SiO maser emission at 
typical distances of  
2--4 stellar radii toward these stars using very long baseline
interferometry (VLBI) at radio wavelengths
\citep[e.g.,][]{BDK:97,KD:97,BM:00,H:01,DK:03}. Dust shells outward of the 
SiO maser emission zone have been studied using mid-infrared 
interferometry \citep[e.g.,][]{D:94,G:95,T:03}. Near- and mid-infrared 
interferometry have also been used to study the warm extended atmosphere of
Mira variables containing gaseous H$_2$O, SiO and CO molecules
\citep[e.g.,][]{PFR:99,MPC:02,OBD:04}.

Theoretical models describe and predict photospheric center-to-limb
intensity variations including the effect of close molecular layers
\citep[e.g.,][]{Bessel:96, HSW:98, Tej:03, ISW:04}, the location and dynamics
of SiO maser emission zones \citep{Humphreys:96,Humphreys:02}, as well as
gas and dust components of the stellar winds
\citep[e.g.,][]{Winters:00,Hoefner:03}.

Multi-wavelength studies of the stellar surface and the CSE, utilizing 
a combination of the above techniques, are well suited to study the structure 
of the CSE, the mass-loss process, and the evolution of these stars.
It is common for infrared observations of the stellar photosphere or the
circumstellar dust to be compared to radio observations of circumstellar 
masers from the literature and vice versa.  For example, \citet{D:94} 
compared the inner radii and extents of dust shells as determined from 
Infrared Spatial Interferometer (ISI) observations with estimates of the 
photospheric radii and measurements of the extents of the SiO, H$_2$O, 
and OH maser shells. Similarly, \cite{MONNIER:04} compared 
high-resolution data on the stellar diameters and dust shells of two stars 
(VX Sgr and NML Cyg) 
from Keck aperture masking and Infrared Optical Telescope Array (IOTA) 
interferometry with previously published maps of the SiO maser emission
toward these stars.  Such comparisons, however, are somewhat limited 
due to the inherent variability in the CSE and the star itself.   
The SiO masers for example have been shown to have proper
motions and intensity variations on time scales of a few weeks 
\citep{BDK:97,DK:03}.  Similarly, interferometric measurements of the 
stellar parameters vary with stellar variability phase as discussed 
above.   This inherent variability of Mira stars and their surroundings
necessitates a more coordinated approach to multi-wavelength studies 
of these objects.
 
The first such coordinated multi-wavelength study of a late-type variable
was performed by \citet{G:95} for the star VX~Sgr.  Coordinated measurements 
included both mid-infrared observations to determine the extent of 
the circumstellar dust shell and VLBI mapping of the SiO maser zone. 
More recently, \citet{C:04} compared Very Long Baseline Array (VLBA) images 
of the SiO maser emission of several Mira stars with near-infrared 
diameters obtained at the Infrared Optical Telescope Array (IOTA) 
interferometer \citep{MPC:02}, and with literature values of the mid-infrared 
dust shell extensions. 

Here, we present the first results from a series of coordinated VLTI/VLBA 
observations of late-type variable stars, in particular, for the Mira variable S~Ori.
Our results include the first radio interferometric images of the SiO masers
(VLBA) and the most accurate photospheric diameter measurements using 
near-infrared $K$-band interferometry (VLTI/VINCI).

\section{CHARACTERISTICS OF S ORI \label{SORI}}

The target of our observations, S~Ori, is a Mira variable star with 
spectral type M6.5-M9.5e \citep{KHOLOPOV:85} and V magnitude 8.4-13.3 
\citep{ASAS}. \citet{WMF:00} report a quite low near-infrared $J$-band 
pulsation amplitude of 0.55\,mag.
They also derive a mean bolometric magnitude $m_\mathrm{bol}=3.08$, 
with an amplitude $\Delta m_\mathrm{bol}=0.49$.  
Over the last 100 years, the variability 
period of S~Ori has been seen to randomly vary
between about 400 and 450 days \citep{Bedding:99, WMF:00, MJ:02}.
For our study, we 
adopt a stellar period of 420 days and a time of last maximum visual 
brightness of JD 2452331, as given by \citet{ASAS}.  This is  
consistent within $\pm\sim$\,5\,days with an inspection 
of the most recent AAVSO and AFOEV data. \citet{BDBL:96} measured the 
near-infrared $K$-band uniform disk (UD) diameter of S~Ori at 
variability phase 0.56 to be 10.54 $\pm$ 0.68 mas, and derived a Rosseland
mean diameter of 11.70 $\pm$ 0.75 mas. We adopt the distance to S~Ori
to be 422 $\pm$ 37 pc, as given by \citet{BDBL:96}, based on
measurements by \cite{WC:83} and \cite{Young:95}. This is consistent with the 
value of 430\,pc from \cite{WMF:00} based on their $K$ period-luminosity 
relation.  S Ori exhibits SiO and OH masers  \citep{Benson:90},
but no detection of any H$_2$O maser emission has been reported.
\cite{SP:98} report a relatively low dust emission coefficient ($DEC$),
i.e. the total emission of the dust to the total emission of the star
in the wavelength range from 7.7--14.0\,$\mu$m, of 0.24 for S Ori
(for comparison, $DEC = 0.46$ for $o$\,Cet).

\section{OBSERVATIONS AND REDUCTION \label{OBS}}

\subsection{Near-infrared $K$-band interferometry \label{sec:obsvlti}}

We obtained near-infrared $K$-band interferometric measurements
of S~Ori using the ESO Very Large Telescope Interferometer (VLTI) 
equipped with the commissioning instrument VINCI and the two VLTI test 
siderostats (effective apertures $\sim$\,40\,cm). 
VLTI stations D\,1 and B\,3, forming an unprojected 
baseline length of 24\,m, were used for 13 nights between 
2003 January 25 (JD 2452665) and  2003 March 31 (JD 2452730).
The S~Ori visual variability phase ranged from 0.80 to 0.95 for these 
dates. Available data taken in 2002 December using an 8\,m ground
baseline and in 2003 September using a 16\,m ground baseline were not used 
because S~Ori was only marginally resolved with these shorter baselines
($|V|\gtrsim 0.9$).   A recent general description of the VLTI can be 
found in \citet{Glindemann:03} and references therein.
During each observation night, several series of typically 500 
interferogram scans were recorded on S~Ori as well as on
several calibration stars.
The calibration stars used and their adopted properties are listed in 
Table~\ref{tab:calibrators}. 
For S~Ori, an effective temperature 
of 2500\,K was used, which is consistent with the value of 
2313 $\pm$ 110 K obtained 
by \citet{BDBL:96} for phase 0.56. We note that variations of the 
effective temperatures up to $\pm$ 500\,K result in variations of
the squared visibility amplitude of less than 0.3\%.
The scan length for the S~Ori and all calibration stars' observations
was 230\,$\mu$m,
the scan speed was 616\,$\mu$m/sec, and the fringe frequency, corresponding
to the time to scan one interferometric fringe, was 289.5\,Hz.

Mean coherence factors were obtained for each series of interferograms
using the VINCI data reduction software, version 3.0, as described by
\citet{Kervella:04}, employing the results based on wavelets transforms.
Calibrated squared visibility values for S~Ori were obtained
by calibration of the mean coherence factors as described in \citet{WAK:04}, 
with a time kernel of 5 hours to convolve the measured transfer function. 
The errors of the calibrated S~Ori squared visibility values include the
scatter of the single-scan coherence factors, the adopted errors of
the diameters of the calibration stars, and the observed variation of the 
transfer function during each night. 
The calibrated visibility
values are listed in Table \ref{tab:visibilities} and are also available
in electronic form from the authors upon request.  Table \ref{tab:visibilities} also
lists the date and time of observation, the spatial frequency,
the position angle of the projected baseline (East of North),
the calibrated squared visibility value $V^2$ and its error $\sigma_{V^2}$,
as well as the number of processed interferograms. The effective
wavelength is 2.19$\mu$m.  All observations were 
performed at similar angles of the projected 
baseline ranging from $72^{\circ}-74^{\circ}$ E of N. 
The squared visibility values for S~Ori were grouped into four bins of 
five nights each (bin width $\sim 1\%$ of the visual variability period
of 420 days).

We characterize our measurements with a best-fitting uniform disk (UD) 
diameter for each of our four epochs. Our uniform disk model takes into 
account the broad passband (1.9--2.5~$\mu$m) of the VINCI instrument, 
as described in \citet{WAK:04}.  
Figure ~\ref{visibilities} shows the obtained squared visibility
amplitudes of S~Ori together with the best fitting uniform disk
models for each of our four VLTI/VINCI epochs. 
The larger scatter of the visibility data for the March 16-21 epoch,
as compared to the other epochs, can be explained by a larger
time difference of the calibration stars. 
Table ~\ref{tab:vinciresults} 
lists the observation dates and the determined $K$-band UD diameters
and their associated formal errors.  Additional calibration uncertainties
are estimated to be $\sim$\,0.1\,mas, based on an analysis of sub-volumes
of the data and differing methods of interpolating the transfer function.
The addition of a possible circumstellar dust shell is not expected to
affect the $K$-band UD diameter significantly
\citep[see for instance discussions in][]{WED:04,OBD:04}.
In particular, \cite{OBD:04} modeled an optically thin
dust disk for a similar Mira star, RR~Sco (dust emission coefficient
0.21, compared to the value of 0.24 for S~Ori), for the purpose of
comparison to mid-infrared interferometric data. At a wavelength of
2.2\,$\mu$m the scattered and thermal emission from this dust disk is
lower by a factor of $\sim$\,1000 compared to the attenuated star light and
the effect on the squared visibility values at $V^2\sim$\,0.5 is less than
0.2\% (Ohnaka, K. 2004, private communication). This would lead to an 
effect on the uniform disk diameters in Tab.~\ref{tab:vinciresults} of less 
than 0.02\,mas.
Discussed in Sect. \ref{sec:photosphere} below is the relationship 
between a $K$-band UD diameter and a physically more meaningful angular 
size of the stellar photosphere.
\subsection{SiO Maser Observations}

We simultaneously observed the $v=1, J=1-0$ (43.1 GHz) and $v=2, J=1-0$ 
(42.8 GHz) SiO maser transitions toward S~Ori
($\alpha = 05^h 29^m 00^{s}.9, \delta = -04^{\circ} 41' 32''.7$, J2000)
on 2002 December 29for 6 hrs starting at 02:43 UT (JD 2452636.6, 
visual variability phase 0.73).  
S~Ori and a continuum calibrator
(0359+509) were observed using the 10 stations of the VLBA.  The VLBA
is operated by the National Radio Astronomy Observatory
(NRAO).\footnote{The National Radio Astronomy Observatory is a
facility of the National Science Foundation operated under cooperative
agreement by Associated Universities, Inc.}   Reference frequencies 
of 43.122080 and 42.820587~GHz were used for the $v=1$ and $v=2$ SiO 
transitions respectively.  Data were recorded in dual circular polarization 
using two 8-MHz (56.1~km~s$^{-1}$) bands centered on the local standard 
of rest (LSR) velocity of 18.0~km~s$^{-1}$.  System temperatures 
and point source sensitivities were on the order of $\sim$150~K 
and $\sim$11~Jy~K$^{-1}$ respectively.

The data were correlated at the VLBA correlator operated by NRAO in 
Socorro, New Mexico.  Auto and cross-correlation
spectra consisting of 256 channels with channel spacings of 31.25~kHz 
($\sim$0.2~km~s$^{-1}$) were produced by the correlator.  Calibration
was performed using the Astronomical Image Processing
System (AIPS) maintained by NRAO.  The total intensity data
were calibrated in accordance with the procedures outlined in
\cite{DIAMOND:89}.  The bandpass response was determined from scans
on the continuum calibrator and used to correct the target source
data. The time-dependent gains of all antennas relative to a reference
antenna were determined by fitting a total-power spectrum (from the
reference antenna with the target source at a high elevation) to the
total power spectrum of each antenna.  The absolute flux density scale
was established by scaling these gains by the system temperature and
gain of the reference antenna.  Errors in the gain and pointing of the
reference antenna and the atmospheric opacity contribute to the error
in the absolute amplitude calibration, which is accurate to about
15--20\%.

To correct any instrumental delay, a fringe fit was performed on
the continuum calibrator scans, and residual group delays for each 
antenna were determined.  Variations in the residual delays ranged
from 2-4~ns resulting in phase errors of no more than 
1.5--3$^{\circ}$ across the 8-MHz band.  Residual fringe-rates were 
obtained by fringe-fitting a strong reference feature in the spectrum
of each maser transition.  For both transitions we used the same channel 
at a velocity $V_{\rm LSR} = 16.2$~km~s$^{-1}$.  The resulting 
fringe-rate solutions were applied to all channels in each spectrum
respectively.  An iterative self-calibration and imaging procedure 
was then performed to map this reference channel for each transition.  
The resulting residual phase and amplitude corrections from the 
reference channels at 42.8 and 43.1 GHz were applied to all channels 
in the respective bands.

In order to accurately compare the distributions of the two maser 
transitions, it is desirable to determine a common spatial reference 
point.  However, after the fringe-fitting step to determine residual 
fringe-rates, all absolute position information is lost for the VLBA data.  
We accomplished the registration of the two transitions by applying the 
fringe-fit solutions from the $v=2$, 42.8-GHz transition to the 
$v=1$, 43.1-GHz data, re-mapping the $v=1$ transition, and comparing 
the resulting images with those produced from the application of the 
$v=1$ calibration itself.  Although the images resulting from the application 
of the 42.8-GHz fringe-fit to the 43.1-GHz data were of poorer quality, 
we were still able to determine an offset between the two sets of images.  
The offsets in right ascension and declination computed from images 
of two different spectral channels, were the same to within 0.03 mas. 
Subsequent phase self-calibration and imaging were performed for the 
$v=1$ SiO data using these computed offsets, resulting in spatially 
aligned $v=1$ and  $v=2$ image cubes.  
This same procedure was applied to the continuum calibrator source,
0359+509, and the positions derived agreed to within 0.4 mas, thus
providing an estimate of the error in the registration of the two maser
distributions.  

Final images of the SiO maser emission consisting of $1024\times 1024$ pixels 
($\sim$$51\times51$~mas) were generated using synthesized beams of 
$0.58\times 0.17$~mas and $0.52\times 0.17$~mas for the $v=1$ and $v=2$
transitions respectively.  Images were produced for 
spectral channels from 4.7~km~s$^{-1}$ to 30.7~km~s$^{-1}$ forming 
image cubes of 120 planes.  Off-source RMS noise in the images ranged from 
7~mJy to 18~mJy.  Figure~\ref{SORI_COMB_IMAX} shows the total intensity 
contour maps of the $v=1$, 43.1-GHz (red) and $v=2$, 42.8-GHz
(blue) SiO maser emission toward S Ori.  The contours represent the 
maximum pixel value in each image cube over the LSR velocity range from 
$+10.1$~km~s$^{-1}$ to $+25.4$~km~s$^{-1}$.  Subsequent analysis of the 
image data is described in detail in Sect.~\ref{MASER_RESULTS}.

\section{RESULTS AND DISCUSSION}

\subsection{The photospheric diameter \label{sec:photosphere}}
 
Figure~\ref{diamdate} and Tab.~\ref{tab:vinciresults} show our  
near-infrared $K$-band uniform disk (UD) diameters of S~Ori determined from
our four VLTI/VINCI epochs as a function of observation date and 
visual variability phase, as discussed in Sect.~\ref{sec:obsvlti}.
The $K$-band UD diameter decreases almost linearly from
$\sim$\,10.5\,mas at visual variability phase 0.80 to $\sim$\,10.2\,mas 
at phase 0.95, i.e. by $\sim$\,3\%. It is not clear whether the local
diameter minimum of $\sim$\,10.0\,mas at phase 0.92 in 
Tab.~\ref{tab:vinciresults} and Fig.~\ref{diamdate} is caused by
an additional systematic calibration error as discussed in 
Sect.~\ref{sec:obsvlti}, or by a real minimum of the stellar size.
The minimum of a Mira $K$-band UD diameter is predicted to lie at
visual variability phase $\sim$\,0.9, according to model predictions
by \cite{ISW:04}.
Our coordinated VLBA measurement at visual variability phase 0.73 occurred
about one month before the first VLTI epoch at phase 0.8, and is hence
close to, but not exactly contemporaneous to the VLTI epochs.
Both, observations of the diameter change of the Mira star 
S\,Lac by \cite{TCB:02}
as well as Mira star model predictions by \cite{ISW:04} show
an almost linear change of the $K$-band UD diameter
between variability phases 0.7 and 0.9, i.e. in the pre-minimum part
of the diameter curve and pre-maximum part of the visual light curve.
Thus, we can linearly extrapolate our $K$-band UD diameter data at phases 
0.8--0.95 to the VLBA phase 0.73.
For phase 0.73, we determine an extrapolated $K$-band UD diameter of
$\Theta_\mathrm{UD}^K (\mathrm{VLBA\,epoch, phase=0.73}) = 10.8 \pm 0.3$\,mas.
This value and its error correspond to the mean and difference of the
two linear extrapolations shown in Fig.~\ref{diamdate}, which used 
all four points and only the two closest points to the VLBA epoch respectively.
This UD diameter corresponds to a linear radius
$R_\mathrm{UD}^K = 2.3 \pm 0.5$~AU or
$R_\mathrm{UD}^K =490 \pm 115~R_\odot$ with our assumed
distance to S~Ori.

A UD intensity profile is often not an ideal
representation of the true near-infrared center-to-limb intensity variation
(CLV) of cool giants in general and of Mira stars in particular. The 
CLV of cool giants has been studied by, e.g. 
\cite{QMB:96,BURNS:97,SCHOLZ:98,HSW:98,WHJ:01,WAK:04,ISW:04,
WED:04}.  UD diameters of non-variable giants
obtained from visibility measurements in the first lobe 
are usually transformed into physically more meaningful Rosseland 
angular diameters using correction factors determined from 
atmosphere models in the literature \citep[e.g.][]{C:03}.  
For the case of Mira variable stars, few atmosphere models
are available.  We used the hydrodynamic atmosphere models from 
\cite{HSW:98,Tej:03,ISW:04},  to estimate the relationship between the 
$K$-band UD diameter and the continuum diameter. While the broad-band
UD diameter is affected by various molecular bands, the continuum diameter
is a better estimate of the real photospheric stellar size. These models
have been constructed for the prototype Mira stars $o$\,Cet and R\,Leo
and have been compared to observations of several Mira stars by,
for instance, \cite{HBSW:01,HOFMANN:02,WED:04}. The general 
model results are not expected to be dramatically different 
for other Mira stars such as S Ori (Scholz, M. 2004, private communication).
The model predicted difference between the continuum and 
$K$-band UD diameters is relatively low in the pre-maximum region of the 
visual variability curve as in the case of our observations.  
 At this phase of 0.73, the continuum diameter may be smaller than 
 the $K$-band UD diameter by about 15\% \citep{ISW:04} .
With this assumption, the continuum photospheric diameter
for the epoch of our VLBA observation would be
$\Theta_\mathrm{Phot}(\mathrm{VLBA\ epoch,\ phase=0.73})\approx 9.2$\,mas.
This angular radius corresponds to a  photospheric radius
of   $R_\mathrm{Phot} \approx 420$\,R$_\odot$
or $R_\mathrm{Phot}\approx 1.9$\,AU with our assumed
distance to S~Ori.
With a bolometric flux of $m_\mathrm{bol}=3.1 \pm 0.3$ \citep{WMF:00},
our value for $\Theta_\mathrm{Phot}$ corresponds to
an effective temperature $T_\mathrm{eff}\approx 2670$\,K.
\subsection{The $v=1$ and $v=2$ SiO Maser Emission \label{MASER_RESULTS}}

Our total intensity images of the 43.1 GHz and 42.8 GHz SiO maser emission 
(Figure~\ref{SORI_COMB_IMAX}) show a typical clumpy distribution of the 
maser spots within a ring-like structure.  For the $v=2$, 42.8-GHz
transition the ring is relatively sparse with nearly all of the masers 
concentrated to the NW side of the shell, and a few features
to the SE.  The $v=1$, 43.1-GHz masers, however, form a more typical 
ring-like structure often seen for other stars with SiO masers. 
Like the $v=2$ masers, the $v=1$ SiO also has a higher concentration of 
features on the NW side of the shell.  The $v=1$ maser ring appears to 
be symmetrical, forming a nearly circular shell.  The distribution of 
the $v=2$ SiO masers is too sparse to allow us to comment on any 
symmetry for this transition.  

In order to identify and parameterize maser components, two-dimensional
Gaussian functions were fit to the emission in each spectral (velocity) 
channel using the AIPS task SAD.  Image quality was assessed using
the off-source RMS noise and the deepest negative pixel in the image.
A cutoff flux density was conservatively 
set to the greater of 8$\sigma_{\rm RMS}$ or the absolute 
value of the deepest negative pixel in the plane.  Features with 
flux densities greater than this cutoff were fit with Gaussians to 
determine component parameters.  Errors in the positions in right
ascension and declination of identified features were computed using
the fitted source size divided by twice the signal-to-noise 
ratio (SNR) in the image and ranged from 1~$\mu$as for features 
with high SNR, to 50~$\mu$as for features with lower SNR. 

Since the $\sim$0.2~km~s$^{-1}$ channel spacing is 
sufficient to resolve the masers spectrally, features typically appear in 
multiple adjacent spectral channels.  Positions in right ascension and  
declination and center velocities for the masers were determined using a 
flux-density-squared weighted average for features identified in two or 
more adjacent channels with a spatial coincidence of 0.2~mas 
($\sim$2/3 of the geometric mean of the synthesized beam).  The flux
assigned to the maser averages was the maximum single-channel
flux density. The maser components identified using this procedure
are represented by the circles in Figures \ref{SORI_43.1_COMPS}
and \ref{SORI_42.8_COMPS}.  In the figures, point sizes are 
proportional to the logarithm of the fitted flux density which ranged 
from 0.2 to 4.3~Jy for Figure \ref{SORI_43.1_COMPS} and 0.1 to 5.3~Jy 
for Figure \ref{SORI_42.8_COMPS}.  The line-of-sight (LOS) velocity 
information 
for the masers is also represented in Figures \ref{SORI_43.1_COMPS}
and \ref{SORI_42.8_COMPS}.  The top panel of each figure shows the 
spectrum of SiO maser emission ranging from 10 to 25~km~s$^{-1}$ 
color-coded by LOS velocity in increments of 2~km~s$^{-1}$.  The bottom 
panels of Figures \ref{SORI_43.1_COMPS} and \ref{SORI_42.8_COMPS} show 
the spatial distribution of the SiO masers plotted with the same velocity 
color-coding as in the top panel, and with the color of the maser representing 
its corresponding velocity range in the spectrum.  Comparing the maser 
component maps with the total intensity contour maps, we see that the 
identified features accurately represent the emission summed over all 
velocity channels in the image cube.  From the component maps there 
does not appear to be any coherent velocity structure indicative 
of global expansion/infall or rotation.  We do note that a group of 
components on the western side of the shell shows a velocity 
gradient with velocity decreasing with increasing distance from the 
star.  Such velocity gradients in the SiO maser emission have been 
observed previously \citep[e.g.][]{BM:00, H:01}.

To compare the size of the SiO maser distribution to the photospheric 
diameter of the star as measured by the near-infrared $K$-band 
interferometry we determined the average distance of the masers from
the center of the distribution.  To accomplish this, we first determined 
the center of the distribution by performing a least-squares fit of a 
circle to the combined $v=1$ and $v=2$ maser component data.  This 
fit produced a common center from which we computed the mean 
maser angular distance $\overline{r}_{\rm SiO}$ and the standard deviation 
for each transition independently.  The mean angular distances from center for 
the observed SiO masers at $v=1$, 43.1-GHz and $v=2$, 42.8-GHz are 
$\overline{r}_\mathrm{SiO}=9.4$ and $\overline{r}_\mathrm{SiO}=8.8$~mas 
respectively.  These distances are indicated by a dashed 
circle and are listed in the lower panel of Figures 
\ref{SORI_43.1_COMPS} and \ref{SORI_42.8_COMPS}.  Standard deviations 
of the distances are $\sigma_\mathrm{SiO}=1.4$ and 
$\sigma_\mathrm{SiO}=1.7$~mas for the $v=1$ and $v=2$ transitions 
respectively.  The standard deviations provide an indication of the 
thickness of the shell and are likely dominated by the features on the 
western side of the shell with a wide range of distances from the center.  
The mean angular distances we derive are consistent with least-squares 
circle fits to each distribution independently using the previously 
mentioned common center.  The least-squares fits to each transition 
have the added assumption that the distribution is circular, thus we 
have chosen to report only the mean distance from center.
The computed angular distances translate to linear shell distances from 
center of $4.0 \pm 0.6$~AU and $3.7 \pm 0.6$~AU at the assumed distance of 
$422 \pm 37$~pc.  The standard deviations are likewise $0.6 \pm 0.1$~AU 
and $0.7 \pm 0.1$~AU for the $v=1$ and $v=2$ masers respectively.

In principal, a comparison of the relative spatial locations of the 
$v=1$ and $v=2$ masers should allow us to comment on possible pumping 
mechanisms (i.e. collisional, radiative, or combination) as in 
\cite{Desmurs:00}.  However, because the $v=2$, 42.8-GHz masers are 
mostly confined to a small region on the NW side of the shell, it is 
difficult to unambiguously determine the relative shift between the 
two maser shells.  Although the mean distance from center we determine
for the $v=2$ masers is slightly smaller that the $v=1$ mean distance,
the two values are consistent to within the errors.

\subsection{Comparison of SiO maser distribution and 
photospheric stellar diameter}

Figures \ref{SORI_43.1_COMPS} and \ref{SORI_42.8_COMPS} also provide a 
comparison of the distribution of the SiO maser spots with our 
obtained angular size of the photospheric disk at the same epoch.
The average distance of the maser spots from the center of their distribution
for the 43.1 GHz and 42.8 GHz transitions 
($\overline{r}_\mathrm{SiO}=9.4$\,mas and $\overline{r}_\mathrm{SiO}=8.8$\,mas)
appears at 1.7\,R$_\star$ and 1.6\,R$_\star$ respectively when compared to 
our estimate of the $K$-band UD diameter.  When compared to our 
estimate of the continuum diameter these values are slightly larger at
2.0\,R$_\star$ and 1.9\,R$_\star$ respectively.  
The widths of the distributions
are approximately 0.3\,R$_\star$ for the masers in both transitions.  
As a reference the angular photospheric radius is listed in the lower 
panel of Figures \ref{SORI_43.1_COMPS} and \ref{SORI_42.8_COMPS} and 
is indicated by a colored circle at the center of the maser distribution.
The position of the near-infrared stellar disk as indicated on the plots is 
assumed to coincide with the center of the SiO maser distribution as
discussed in Sect.~\ref{MASER_RESULTS}.  The true location of the 
star relative to the masers is still unknown.
We note that the above estimates are based on angular sizes derived for the SiO
maser emission and the near-infrared photosphere, and hence are independent 
of the distance to S Ori.  This result is consistent with theoretical
estimates by \cite{Humphreys:02} based on a stellar hydrodynamic pulsation
model combined with an SiO maser model.  They obtain values 
between 1.7\,R$_\star$ and 2.1\,R$_\star$, depending on the variability phase.

Our result is consistent with a SiO maser ring radius of approximately 2\,R$_\star$ 
as determined by \cite{C:04} for several Mira variables.  Recently, however, \cite{MONNIER:04} 
updated the stellar diameter estimate of \cite{G:95} for the supergiant VX~Sgr and 
found that the SiO masers lie at a greater distance from the photosphere, 3.9\,R$_\star$, 
rather than the 1.3\,R$_\star$ determined by \cite{G:95}.  It is striking that the stellar
diameter for the other supergiant observed by \cite{MONNIER:04}, NML Cyg,  also indicates
a SiO maser ring distance of $\sim$4\,R$_\star$ when compared to the 
SiO maser shell diameter obtained by \cite{BM:00}.
It is unclear whether
these larger distances are an artifact caused by the non-contemporaneous measurements or, 
whether they indicate an inherent difference between Miras and supergiants.
In earlier studies, measured sizes for SiO maser shells have been compared 
to stellar sizes based on diameters found in the literature \cite[e.g.][]{BDK:97,BM:00,H:01}.
These diameters are often widely spaced in time and stellar phase from the 
measurements of the SiO maser shell.  
Sometimes the stellar diameters have not been measured at all and comparisons are 
made with even less precise diameter estimates based on the luminosity 
of the star \cite[e.g.][]{DIAMOND:94,COLOMER:96,SANCHEZ:02}.
These examples demonstrate that it is desirable to compare SiO maser ring
diameters with stellar diameter estimates, and highlight the need for contemporaneous
multi-wavelength observations.  

\section{SUMMARY}

Using the VLBA and VLTI/VINCI, we have undertaken a coordinated 
multi-wavelength study of the Mira variable S Ori.   The VLBA observations 
resulted in the first interferometric images of the $v=1,J=1-0$, 43.1-GHz 
and the $v=2, J=1-0$, 42.8-GHz SiO transitions.  These images  
show that the masers lie in a clumpy ring-like distribution.  Analysis of 
the VLBA images provided mean distances from center of 9.4~mas (4.0~AU) 
and 8.8~mas (3.7~AU) and standard deviations of 1.4~mas (0.6~AU) 
and 1.7~mas (0.7~AU) for the $v=1$ and $v=2$ SiO masers respectively 
at a stellar phase of 0.73.  
From near-infrared $K$-band visibility measurements made with VLTI/VINCI, 
we derived UD diameters that decreased over time from $\sim$\,10.5\,mas 
at stellar phase 0.80 to $\sim$10.2\,mas at phase 0.95.   After extrapolating 
the UD diameter to the phase of our VLBA observations and considering
a correction from UD diameter to continuum photospheric size, we 
determine a photospheric 
diameter of  $\Theta_\mathrm{Phot} \approx 9.2$\,mas, corresponding to 
a linear radius of $R_\mathrm{Phot} \approx 1.9$\,AU.  

Because these observations of S Ori were closely spaced in time in a 
coordinated effort between the two instruments, we are able to relate the 
stellar diameter with the size of the SiO maser shell without the uncertainty
caused by the inherent variability of the star.   Thus we can conclusively 
say that the SiO masers lie relatively close to the stellar photosphere at a distance 
of $\sim$2\,R$_\star$ at the time of our observations.

The observations presented here represent the first in a series of 
coordinated VLBA/VLTI experiments to study long-period variable stars.  
The present VLTI observations were obtained during the
commissioning period of the VLTI with the commissioning instrument
VINCI and the small (40\,cm) test siderostats. Future observations will 
utilize the scientific VLTI instruments in the mid-infrared (MIDI) and 
near-infrared (AMBER) with the 8~m Unit Telescopes and 1.8~m 
Auxiliary Telescopes.  The upcoming near-infrared VLTI instrument AMBER 
will allow interferometric measurements in the near-infrared $J$, $H$, and $K$
bands with a spectral resolution of up to 10\,000, and will
provide closure phases. This will enable us to directly measure the
the continuum stellar diameter
and to study the conditions on the stellar surface,
including possible asymmetries and surface inhomogeneities.

\acknowledgements
The near-infrared results are based on public data collected at the 
ESO VLTI, Paranal, Chile, in the framework of our P70 shared
risk program.  We acknowledge support by the ESO DGDF. 
We are grateful to M.~Scholz for valuable comments with respect
to definitions of Mira star radii. We thank as well T.~Driebe and
K.~Ohnaka for helpful discussions.
MW is grateful for hospitality at the U.S. Naval Observatory.
DAB is grateful for the hospitality at the European Southern Observatory.

\clearpage

\begin{deluxetable}{llll}
\tablecaption{Properties of the observed VLTI/VINCI calibration stars
\citep{Borde:02}. \label{tab:calibrators}}
\tablewidth{0pt}
\tablehead{  & \colhead{$\Theta_\mathrm{UD}^K$} & 
\colhead{$\sigma(\Theta)$} &
\colhead{$T_\mathrm{eff}$} \\
\colhead{Star} & \colhead{(mas)} & \colhead{(mas)} &
\colhead{(K)}
}
\startdata
 18 Mon	 	&1.86   & 0.023  & 4656 \\
 31 Ori        	&3.56   & 0.057  & 4046 \\
 58 Hya        	&3.12   & 0.035  & 4318 \\
 $\alpha$ CMa	&5.94   & 0.016  & 9900 \\
 $\delta$ Lep	&2.56   & 0.041  & 4656 \\
 $\epsilon$ Lep	&5.91   & 0.064  & 4046 \\
 HD 112213	&3.15   & 0.036  & 3690 \\
 HD 132833	&3.06   & 0.034  & 3690 \\
 HR 2305	&1.76   & 0.031  & 4256 \\
 HR 2311	&2.43   & 0.040  & 4046 \\
 HR 3803	&6.93   & 0.079  & 4046 \\
 $\xi^2$ Sgr	&3.28   & 0.036  & 4508 \\
 $\nu^2$ CMa	&2.38   & 0.026  & 4497 \\
 $\tau$ Sgr	&3.83   & 0.043  & 4444 \\
\enddata
\end{deluxetable}

\clearpage

\begin{deluxetable}{llccrrr}
\tablecaption{Calibrated VLTI/VINCI visibility values.  \label{tab:visibilities}}
\tablewidth{0pt}
\tablehead{
\colhead{} & \colhead{} & \colhead{Spat. Freq.} & \colhead{Basel. Angle} & 
\colhead{ } & \colhead{ } & \colhead{ } \\
\colhead{Date} & \colhead{Time} & \colhead{($1/''$)} & \colhead{($^\circ$ E of N)} & 
\colhead{$V^2$} & \colhead{$\sigma_{V^2}$} & \colhead{\#}}
\startdata
2003-01-26&01:56:48& 52.88 &  72.23 &4.447e-01 &1.053e-02 & 479\\
2003-01-26&02:02:42& 52.98 &  72.38 &4.431e-01 &1.056e-02 & 458\\
2003-01-26&02:09:00& 53.05 &  72.53 &4.358e-01 &1.062e-02 & 416\\
2003-01-26&02:47:04& 52.68 &  73.19 &4.465e-01 &1.093e-02 & 453\\
2003-01-26&02:53:21& 52.49 &  73.26 &4.462e-01 &1.101e-02 & 436\\
2003-01-26&02:59:39& 52.26 &  73.31 &4.455e-01 &1.106e-02 & 462\\
2003-01-26&03:11:06& 51.76 &  73.38 &4.689e-01 &1.194e-02 & 432\\
2003-01-26&03:17:07& 51.45 &  73.40 &4.576e-01 &1.208e-02 & 412\\
2003-01-26&03:23:36& 51.08 &  73.41 &4.639e-01 &1.495e-02 & 377\\
2003-01-26&04:11:15& 47.21 &  73.03 &5.237e-01 &2.989e-02 & 256\\
2003-01-26&04:17:31& 46.56 &  72.92 &5.417e-01 &3.260e-02 & 250\\
2003-01-26&04:36:44& 44.37 &  72.45 &5.615e-01 &2.009e-02 & 241\\
2003-01-26&04:43:26& 43.55 &  72.25 &5.695e-01 &1.948e-02 & 210\\
2003-01-31&02:40:32& 52.24 &  73.32 &4.838e-01 &4.282e-02 & 145\\
2003-01-31&02:51:39& 51.75 &  73.38 &4.662e-01 &4.361e-03 & 479\\
2003-01-31&02:57:52& 51.43 &  73.40 &4.753e-01 &4.382e-03 & 485\\
2003-01-31&03:49:26& 47.43 &  73.07 &5.284e-01 &7.397e-03 & 365\\
2003-02-05&03:53:23& 44.86 &  72.56 &5.598e-01 &1.807e-02 & 364\\
2003-02-05&04:05:16& 43.40 &  72.21 &5.980e-01 &2.105e-02 & 232\\
2003-02-06&03:26:20& 47.38 &  73.06 &5.323e-01 &1.217e-02 & 471\\
2003-02-06&03:58:48& 43.72 &  72.29 &5.949e-01 &1.483e-02 & 388\\
2003-02-10&01:08:45& 53.04 &  72.50 &5.110e-01 &4.295e-02 & 159\\
2003-02-10&01:28:27& 53.03 &  72.91 &4.723e-01 &2.416e-02 & 427\\
2003-02-10&02:17:13& 51.50 &  73.40 &4.962e-01 &2.543e-02 & 454\\
2003-02-10&02:23:25& 51.15 &  73.41 &4.950e-01 &2.536e-02 & 456\\
2003-02-10&02:29:57& 50.74 &  73.41 &5.159e-01 &2.668e-02 & 443\\
2003-02-10&03:24:02& 45.96 &  72.80 &5.803e-01 &4.593e-02 & 250\\
2003-03-17&00:26:00& 49.77 &  73.36 &5.115e-01 &7.606e-03 & 323\\
2003-03-17&00:33:21& 49.18 &  73.30 &5.121e-01 &1.296e-02 & 245\\
2003-03-17&00:42:06& 48.42 &  73.21 &5.156e-01 &1.454e-02 & 192\\
2003-03-18&00:12:28& 50.47 &  73.40 &5.041e-01 &7.892e-03 & 459\\
2003-03-18&00:25:52& 49.47 &  73.33 &5.173e-01 &9.933e-03 & 309\\
2003-03-18&00:33:16& 48.85 &  73.27 &5.271e-01 &1.035e-02 & 345\\
2003-03-21&00:43:22& 46.75 &  72.95 &5.772e-01 &9.288e-03 & 385\\
2003-03-21&00:48:38& 46.19 &  72.85 &6.055e-01 &2.220e-02 &  42\\
2003-03-21&01:06:55& 44.07 &  72.38 &6.018e-01 &1.650e-02 & 323\\
2003-03-22&00:34:15& 47.29 &  73.04 &5.797e-01 &4.487e-03 & 454\\
2003-03-22&00:41:19& 46.55 &  72.92 &5.837e-01 &4.348e-03 & 471\\
2003-03-22&00:48:31& 45.77 &  72.76 &5.960e-01 &4.632e-03 & 461\\
2003-03-27&23:51:52& 49.03 &  73.29 &5.253e-01 &1.073e-02 & 460\\
2003-03-27&23:59:06& 48.39 &  73.21 &5.368e-01 &1.139e-02 & 389\\
2003-03-28&00:06:22& 47.71 &  73.11 &5.500e-01 &1.248e-02 & 322\\
2003-03-28&23:43:31& 49.40 &  73.32 &5.304e-01 &1.179e-02 & 400\\
2003-03-28&23:50:53& 48.78 &  73.26 &5.279e-01 &1.467e-02 & 348\\
2003-03-28&23:58:29& 48.09 &  73.17 &5.558e-01 &2.190e-02 & 272\\
2003-03-29&23:34:06& 49.83 &  73.36 &5.201e-01 &9.993e-03 & 447\\
2003-03-29&23:41:37& 49.23 &  73.31 &5.238e-01 &1.175e-02 & 352\\
2003-03-29&23:49:13& 48.58 &  73.23 &5.424e-01 &1.325e-02 & 335\\
2003-03-30&00:31:28& 44.08 &  72.38 &6.009e-01 &1.131e-02 & 459\\
2003-03-31&23:37:44& 48.89 &  73.27 &5.267e-01 &2.318e-02 & 342\\
2003-04-01&00:03:01& 46.45 &  72.89 &5.604e-01 &1.989e-02 & 402\\
\enddata
\end{deluxetable}

\clearpage

\begin{deluxetable}{llccrcr}
\tablecaption{VLTI/VINCI observations and derived 
$K$-band uniform disk (UD) diameters with associated 
formal errors. \label{tab:vinciresults}}
\tablewidth{0pt}
\tablehead{
\colhead{} & \colhead{Mean}& \# of 
&  \colhead{Mean} &  
\colhead{$\Theta_\mathrm{UD}^K$} & 
\colhead{$\sigma(\Theta)$} & \colhead{} \\
\colhead{Date} & \colhead{JD}& series 
&  \colhead{Phase} &  
\colhead{(mas)} &  \colhead{(mas)} & \colhead{$\chi^2_\nu$}
}
\startdata
Jan. 25 - 30 & 2452666.8 & 17 & 0.80 & 10.52 & 0.03 & 0.50 \\
Feb. 04 - 09 & 2452678.8 &  10 & 0.83 &  10.33 & 0.07 &  0.55 \\
Mar. 16 - 21& 2452718.0 & 12 & 0.92 &  9.98 & 0.06 &  3.91 \\
Mar. 27 - 31 & 2452728.1 & 12 & 0.95 &  10.16 &  0.03 &  0.13 \\
\enddata
\tablecomments{The observation dates were 
grouped into 4 bins of 5 days each.}
\end{deluxetable}

\clearpage

\centering{Figure Captions}

\figcaption{Near-infrared $K$-band squared visibility amplitudes
of S~Ori and best fitting uniform disk models for our
four VLTI/VINCI epochs. \label{visibilities}}
\figcaption{Total intensity image of the $v=1,J=1-0$ (red) and 
$v=2,J=1-0$ (blue) SiO maser emission toward S~Ori.  The images 
represent maximum pixel values over the LSR velocity range from 
$+10.1$~km~s$^{-1}$ to $+25.4$~km~s$^{-1}$ plotted as contours.  
Contour levels are 1, 2, 4, 8, 16, 32, 64 times the 3$\sigma$ off 
source noise of 0.1~mJy~beam$^{-1}$.  
Peak flux densitiy is 4.36~Jy~beam$^{-1}$ for the $v=1,J=1-0$ line and 
5.35~Jy~beam$^{-1}$ for the $v=2,J=1-0$ transition.  Synthesized beam
sizes are $0.58\times 0.17$~mas at a position angle 
of -15.4$^{\circ}$ for $v=1,J=1-0$ and $0.52\times 0.17$~mas at a 
position angle of -15.7$^{\circ}$ for $v=2,J=1-0$. \label{SORI_COMB_IMAX}}
\figcaption{Obtained S~Ori $K$-band uniform disk diameter as a function 
of date and variability phase. The dashed line denotes the best fitting
linear functions to all of our values and to the two closest points 
to our VLBA epoch.  Our VLBA and VLTI variability phases are indicated by 
arrows. \label{diamdate}}
\figcaption{LOS velocity structure of the $v=1,J=1-0$ SiO maser emission 
toward S~Ori.  The top panel shows the spectrum formed 
by plotting maser intensity versus velocity, color coded in 
2~km~s$^{-1}$ velocity increments from redward (left) to blueward
(right).  The solid line in the top panel represents the scalar-averaged 
cross-power spectrum averaged over all of the VLBA antennas.
The bottom panel plots the spatial and velocity 
distribution of the masers.  The color of each point represents 
the corresponding velocity bin in the spectrum and the size of each 
point is proportional to the logarithm of the flux density.  
Errors in the positions of the features are smaller than the data 
points. The dashed circle is based on the mean angular distance of 
the SiO masers from the center of the distribution.  The colored
circle in the center shows the angular size of the photosphere 
as determined from our VLTI $K$-band measurements. \label{SORI_43.1_COMPS}}

\figcaption{LOS velocity structure of the $v=2,J=1-0$ SiO maser emission 
toward S~Ori.  The top panel shows the spectrum formed 
by plotting maser intensity versus velocity, color coded in 
2~km~s$^{-1}$ velocity increments from redward (left) to blueward
(right).  The solid line in the top panel represents the scalar-averaged 
cross-power spectrum averaged over all of the VLBA antennas.
The bottom panel plots the spatial and velocity 
distribution of the masers.  The color of each point represents 
the corresponding velocity bin in the spectrum and the size of each 
point is proportional to the logarithm of the flux density.  
Errors in the positions of the features are smaller than the data 
points. The dashed circle is based on the mean angular distance of 
the SiO masers from the center of the distribution.  The colored
circle in the center shows the angular size of the photosphere 
as determined from our VLTI $K$-band measurements. \label{SORI_42.8_COMPS}}
\clearpage
\begin{figure}[hbt]
\epsscale{1.0}
\plotone{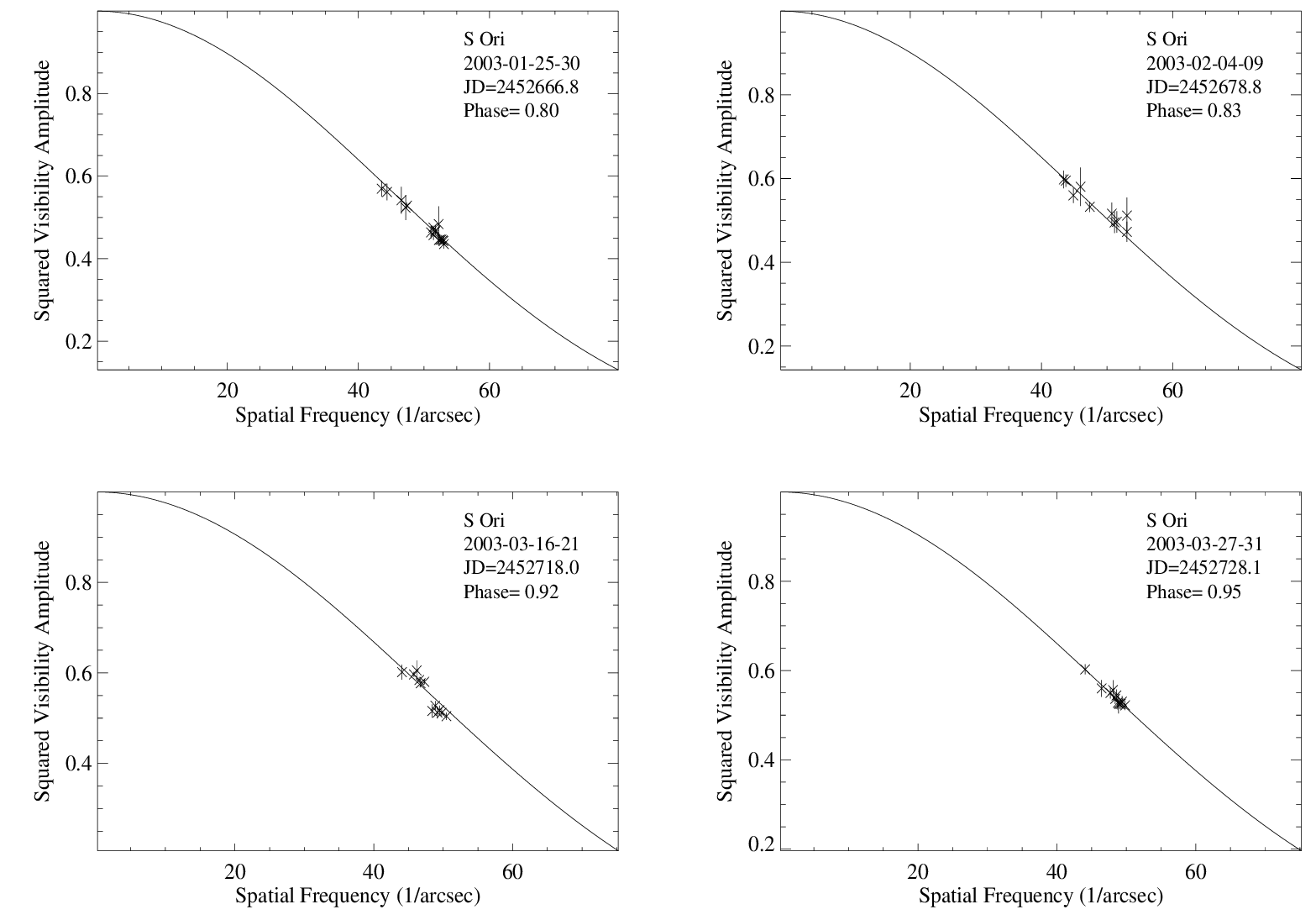}
\centerline{Figure \ref{visibilities}}
\end{figure}
\begin{figure}[hbt]
\plotone{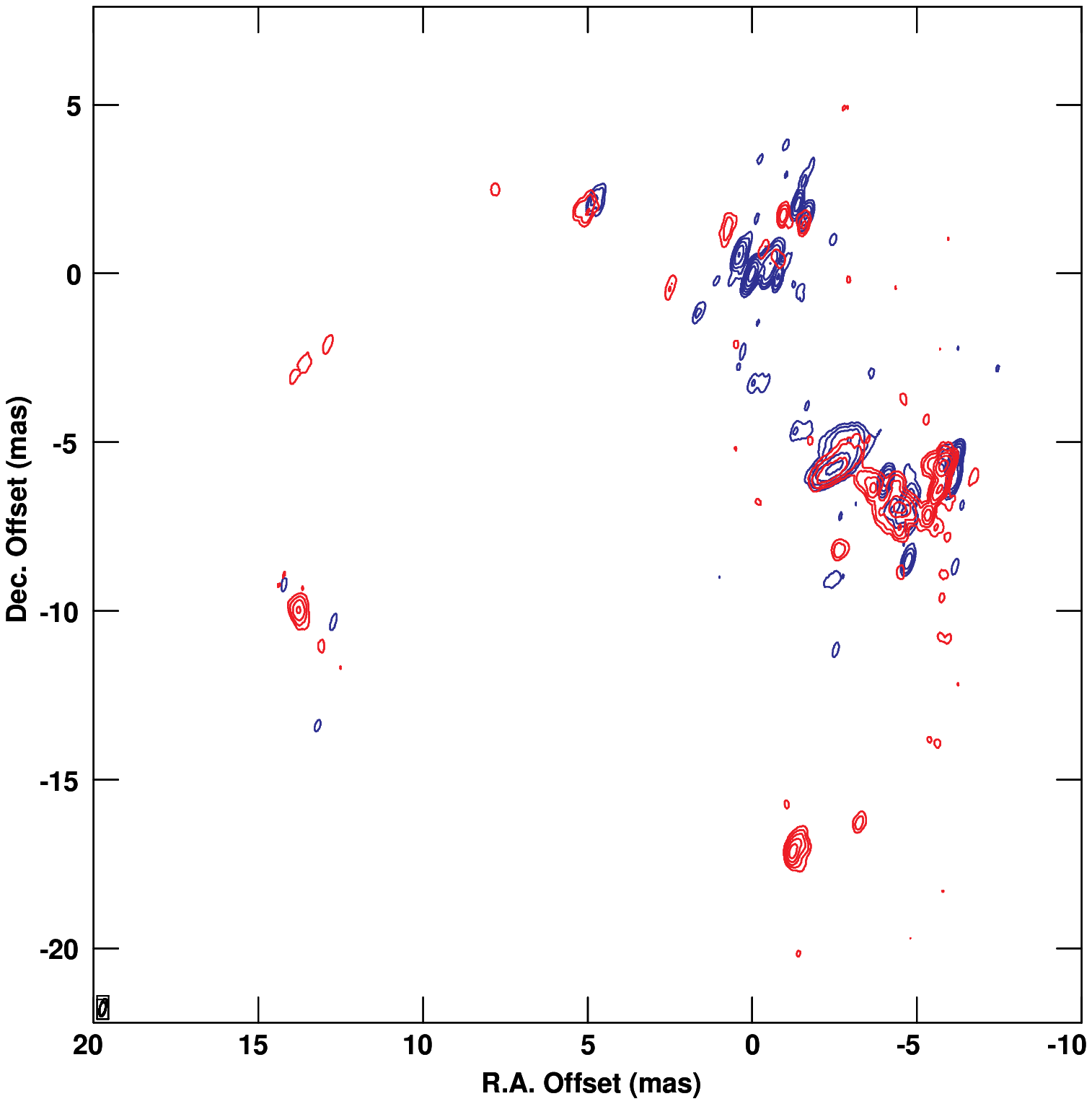}
\centerline{Figure \ref{SORI_COMB_IMAX}}
\end{figure}
\begin{figure}[hbt]
\epsscale{1.0}
\plotone{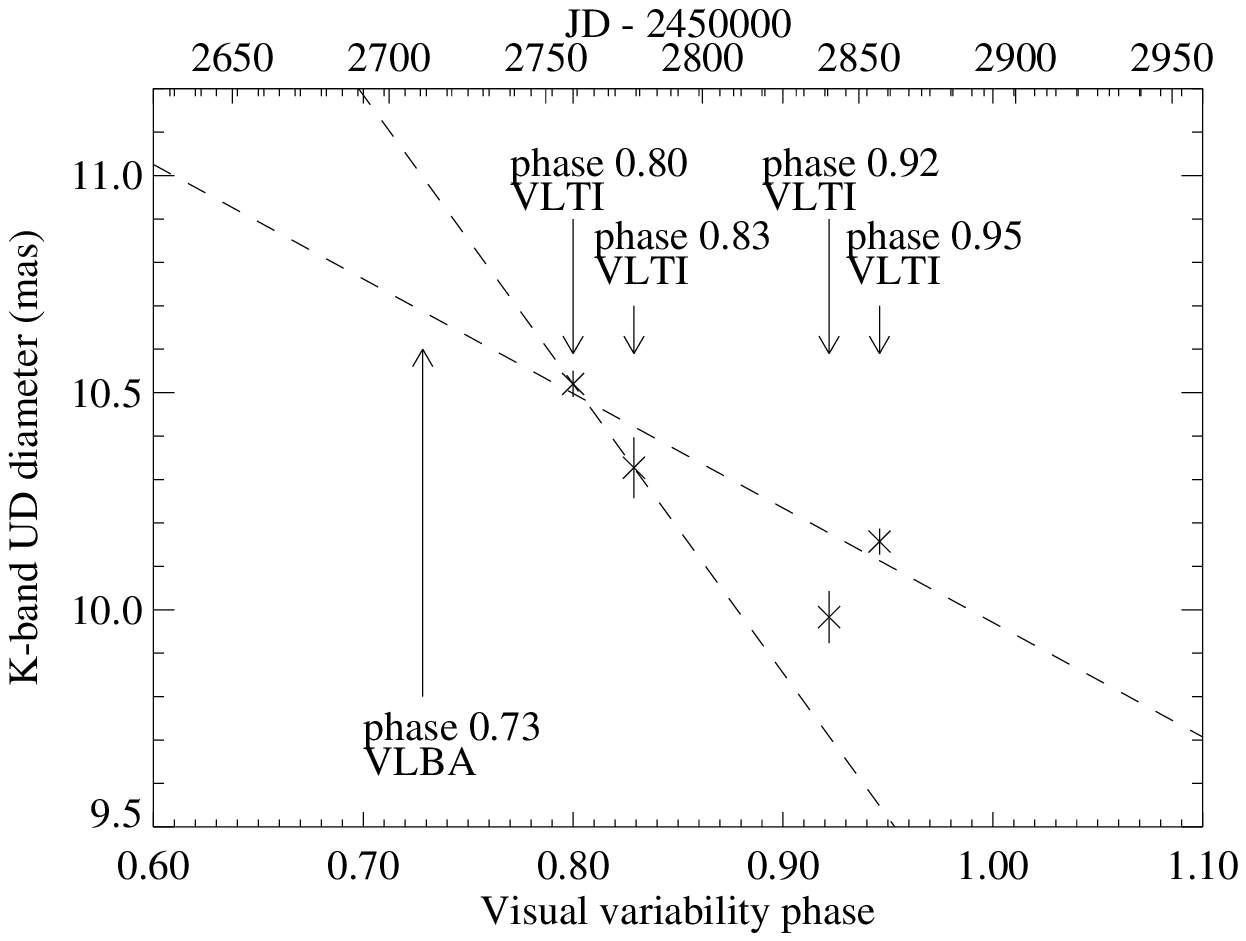}
\centerline{Figure \ref{diamdate}}
\end{figure}
\epsscale{1.0}
\begin{figure}[hbt]
\plotone{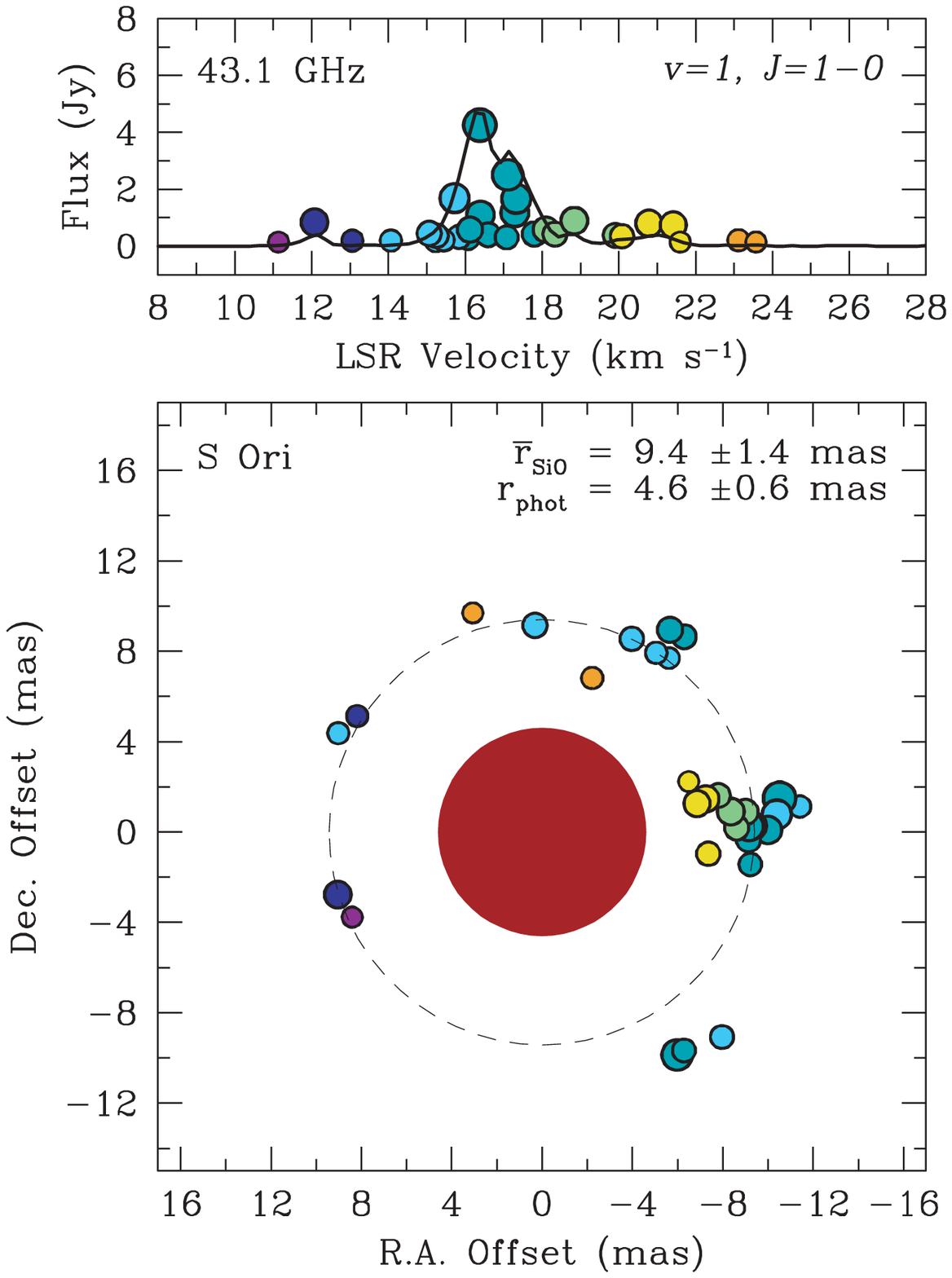}
\centerline{Figure \ref{SORI_43.1_COMPS}}
\end{figure}
\begin{figure}[hbt]
\plotone{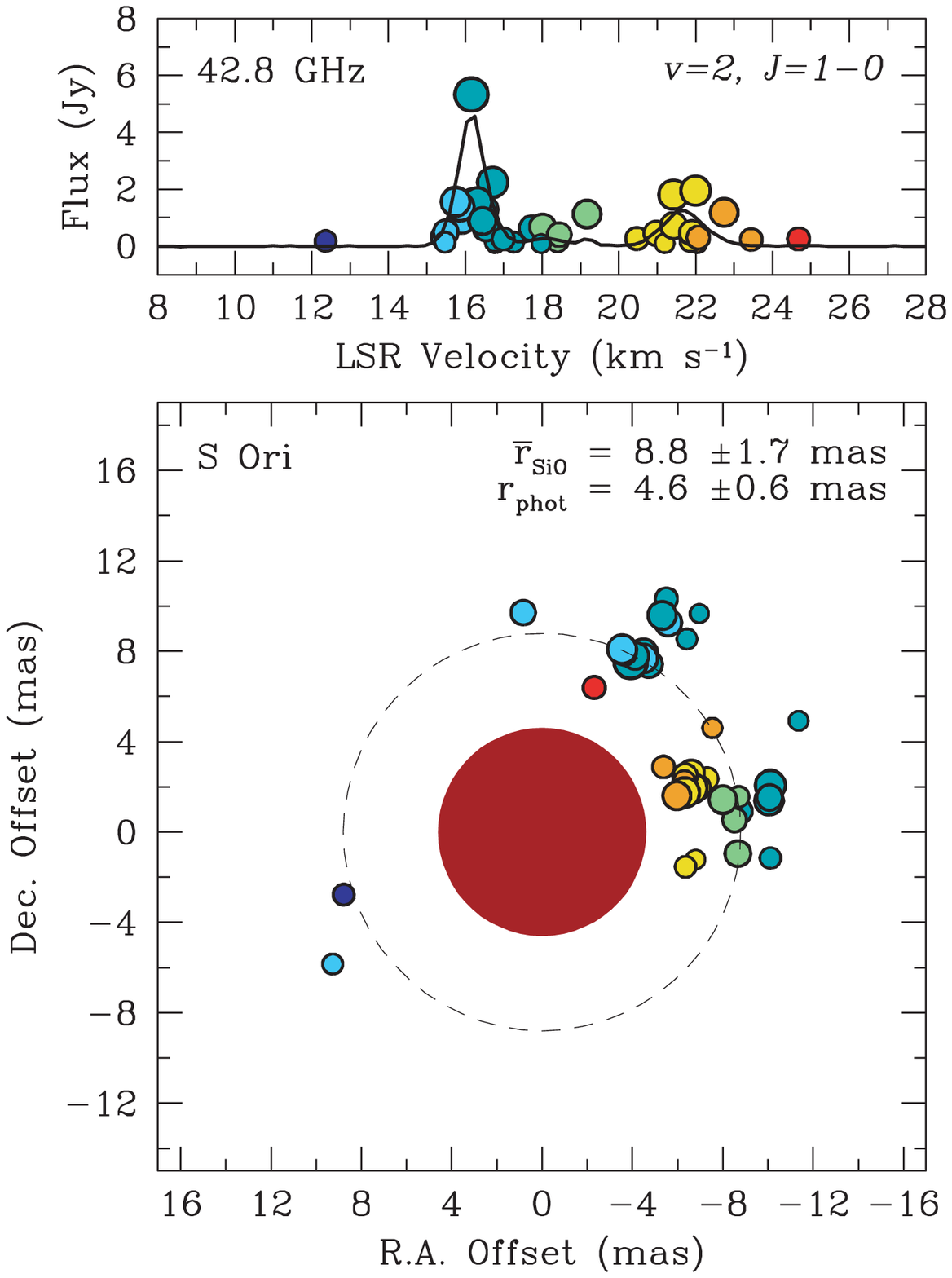}
\centerline{Figure \ref{SORI_42.8_COMPS}}
\end{figure}
\end{document}